\documentclass[11pt,a4paper]{article}
\usepackage{jheppub_kim}
\usepackage{rotating} 
\usepackage{graphicx,epsfig}
\usepackage{amsmath}
\usepackage {amssymb}
\usepackage{subfigure}

\usepackage{relsize}

\newcommand{\be}{\begin{eqnarray}}
\newcommand{\ee}{\end{eqnarray}}
\newcommand{\bea}{\begin{eqnarray}}
\newcommand{\nn}{\nonumber}
\newcommand{\eea}{\end{eqnarray}}

\def\a{\alpha}
\def\b{\beta}
\def\g{\gamma}

\def\d{\delta}

\def\la{\lambda}

\def\m{\mu}
\def\n{\nu}

\def\f{\phi}

\begin{document}

\title{Bi-scalar modified gravity and cosmology with conformal invariance}

\author[a,b]{Emmanuel N. Saridakis}

\author[c]{Minas Tsoukalas}

\affiliation[a]{CASPER, Physics Department, Baylor University, Waco, TX 76798-7310, USA}
\affiliation[b]{Instituto de F\'{\i}sica, Pontificia
Universidad de Cat\'olica de Valpara\'{\i}so, Casilla 4950,
Valpara\'{\i}so, Chile}
\affiliation[c]{Physics Department, Bo\u{g}azi\c{c}i University, 
34342, Bebek, Istanbul, Turkey}

\emailAdd{Emmanuel$_-$Saridakis@baylor.edu}

\emailAdd{minasts@central.ntua.gr}

\abstract{We investigate the cosmological applications of a bi-scalar modified gravity
that exhibits partial conformal invariance, which could become full conformal 
invariance in the absence of the usual Einstein-Hilbert term and introducing additionally 
either the Weyl derivative or properly rescaled fields. Such a theory is constructed by 
considering the action of a non-minimally conformally-coupled scalar field, and adding a 
second scalar allowing for a  nonminimal derivative coupling  with the Einstein tensor 
and 
the energy-momentum tensor of the first field. At a cosmological framework we obtain an 
effective dark-energy sector constituted from both scalars. In the absence of an 
explicit matter sector we extract analytical solutions, which for some parameter regions
correspond to an effective matter era and/or to an effective radiation era, thus the two 
scalars give rise to ``mimetic dark matter'' or to ``dark radiation'' respectively.
In the case where an explicit matter sector is included we obtain a cosmological 
evolution in agreement with observations, that is a transition from matter to dark energy 
era, with the onset of cosmic acceleration. Furthermore, for particular parameter 
regions, 
the effective dark-energy  equation of state can transit to the phantom regime at late 
times. These behaviors reveal the capabilities of the theory, since they arise purely 
from the novel, bi-scalar construction and the involved couplings between the two fields.
}


\maketitle

\section{Introduction}

The motivation for a gravitational modification, i.e the construction of a modified 
theory of
gravity  that possesses General Relativity as a particular limit is twofold. 
On one hand, from purely theoretical considerations, such a modification could improve 
the renormalizability issues of General Relativity, possibly opening the way towards 
gravitational quantization \cite{Joyce:2014kja}. On the other hand, one has the 
cosmological motivation, namely the hope that a gravitational modification could describe 
the observed late-time universe acceleration and/or the early-time inflationary era
\cite{Clifton:2011jh}, without the need of dark energy \cite{Copeland:2006wr,Cai:2009zp} 
or of the inflaton field \cite{Martin:2013tda}.

One of the main requirements when modifying gravity is the preservation of second-order 
field equations, which protects the theory from ghost instabilities  
\cite{ostro,Woodard:2006nt,Woodard:2015zca}. A nice paradigm  along this 
direction are the scalar-tensor theories, which provide one of the simplest 
ways to deviate from General Relativity, and have long been extensively investigated 
\cite{Fujii:2003pa}. Recently, they have regained a significant amount of attention, with 
the resurrection \cite{Charmousis:2011bf,Kobayashi:2011nu} and rediscovery 
\cite{Deffayet:2011gz} of the most general scalar tensor 
theory in four dimensions, with second-order field equations for the metric and the 
scalar field, presented originally in the $70$'s by Horndeski \cite{Horndeski:1974wa}. 
Additionally, lately there have also been attempts to go beyond Horndeski's original 
theory, in terms of the allowed number of derivatives in the field equations, and allow 
for higher derivative theories which nevertheless can be 
cast into second-order form 
\cite{Gleyzes:2014dya,Gleyzes:2014qga,Gao:2014soa,Zumalacarregui:2013pma}. In these 
theories the degrees of freedom remain $2+1$ regardless of the chosen gauge 
\cite{Domenech:2015tca,Langlois:2015cwa,Deffayet:2015qwa,Langlois:2015skt,
Crisostomi:2016tcp}, and the existence of a specific primary constraint allows 
for such theories to survive \cite{Crisostomi:2016czh}. 

Following the above lines, one could try to modify gravity by adding more extra degrees 
of freedom to General Relativity, expressing them as additional scalars, resulting to 
bi-scalar gravitational theories, possessing 2+2 ghost-free degrees of freedom. Such 
bi-scalar theories were originally formulated through extensions of Galileon theories 
\cite{Nicolis:2008in,Deffayet:2009wt,Deffayet:2009mn}, including one extra scalar field 
\cite{Padilla:2010de}. It was later conjectured that the most general covariant 
multi-scalar-tensor theory arises as a natural generalization of Horndeski's theory 
\cite{Padilla:2012dx}, nevertheless it was shown that the proposed theory is 
actually the most general in a flat background, but not in any curved geometry 
\cite{Sivanesan:2013tba,Kobayashi:2013ina}. The most general second-order field equations 
for a bi-scalar theory where recently presented in \cite{Ohashi:2015fma}, 
while in 
\cite{Naruko:2015zze} the authors constructed the Jordan-frame version of general classes 
of bi-scalar theories, which prove to have interesting cosmological implications 
\cite{Saridakis:2016ahq}. Moreover, bi-scalar theories can have interesting 
phenomenology \cite{Padilla:2010tj}. Finally, examining the behavior of these theories 
under conformal transformations could help to reveal their underlying properties
\cite{Padilla:2013jza}.

On the other hand, it is know that conformal invariance is an important property of a 
theory, both theoretically as well as concerning its applications. In particular, 
a gravitational modification with full or partial conformal invariance 
has the theoretical interest that it could be a useful tool towards the exploration of 
physics close to the Planck scale \cite{'THooft:2015skl}. Additionally, it
can have important cosmological implications since it can naturally 
lead to (almost) scale invariant spectrum of primordial density fluctuations, in 
agreement with observations \cite{Ade:2015xua,Sievers:2013ica}. Moreover, 
scale/conformal invariance appears to be an important ingredient in early universe 
cosmology \cite{Bars1,Bars:2011aa,Bars:2012mt,Bars:2012fq,Bars:2013yba,Kallosh}, giving 
rise to a wide range of inflationary models \cite{Kallosh:2013hoa}. Furthermore, it can 
also lead to interesting black hole physics  \cite{thooft}. Finally, there have been 
numerous attempts to explore conformal invariance through Weyl gravity 
\cite{Mannheim:1988dj,Flanagan:2006ra} and its connection  with General Relativity 
\cite{Maldacena:2011mk}.

A bi-scalar theory exhibiting partial conformal invariance was constructed in 
\cite{Charmousis:2014zaa}, which can exhibit full conformal invariance in the absence 
of the usual Einstein-Hilbert term in the action,  introducing additionally either the 
Weyl derivative or properly rescaled fields \cite{Charmousis:2014zaa,Padilla:2013jza}.
Such a construction corresponds to a higher-order version of the usual conformal 
coupling, 
first studied by Bocharova, Bronnikov and Melnikov in  \cite{BBMB1}  and independently 
investigated by Bekenstein in \cite{BBMB2}, namely the BBMB action (see also 
\cite{Ricardo,Martinez:2005di,Martinez:2006an,Martinez:2002ru}). In this theory, one 
considers an additional Galileon field, allowing for a nonminimal derivative coupling  
\cite{Saridakis:2010mf}
with the Einstein tensor and the energy momentum tensor of the first, conformally coupled 
scalar field \cite{Deser:1970zza}, without a potential i.e. maintaining the shift 
symmetry of the additional scalar. The black-hole application of this theory leads to the 
interesting result of 
the presence of a primary scalar   hair  \cite{Charmousis:2014zaa} (see 
\cite{Ogawa:2015pea} regarding possible black-hole instability issues  in shift-symmetric 
theories).

In this work we are interested in investigating the cosmological implications of the 
above bi-scalar theory with (partial) conformal invariance. In particular, we desire to 
extract analytical solutions and study the late-time evolution of a universe 
governed by such a gravitational modification. The plan of the manuscript is the 
following: In Section \ref{Model} we present the construction of the theory, while in 
Section \ref{Cosmolappl} we apply it in a cosmological framework, giving the relevant 
equations. Then, in Section \ref{SolutionsSec} we extract various analytical solutions, 
as well as we perform a numerical elaboration of the full cosmological system. Finally, 
in \ref{Conclusions} we summarize the obtained results.

\section{Action and field equations}
\label{Model}

In this section we demonstrate how to gradually build the action of the particular 
bi-scalar 
theory, which parts of it are conformally invariant, and then we extract the general 
field 
equations. For more information on the procedure, one can look at 
\cite{Charmousis:2014zaa}. We 
start from the usual conformally-coupled action, often called  
Bocharova-Bronnikov-Melnikov-Bekenstein (BBMB) action
\cite{BBMB1,BBMB2} 
which we denote it as $S_{0}$, since it is going to represent the ``seed'' action, namely 
\be
\label{bek}
S_0=\int dx^{4} \sqrt{-g}\; \left\{\frac{1}{16\pi G} R+\eta 
\left[-\frac{1}{2}(\partial\phi)^{2}-\frac{1}{12}\phi^{
2}R\right]\right\},
\ee 
where $G$ is the Newton's constant. In the above action apart from the 
Einstein Hilbert 
term, an additional scalar field $\phi$, conformally coupled to the Ricci scalar $R$, has 
also been 
considered, and we 
have introduced the dimension-less coupling parameter $\eta$. As one can easily see the 
term in the brackets is invariant under the transformation   \cite{Deser:1970zza}
\begin{eqnarray}
&&g_{\m\n}\rightarrow 
\Omega(x)^{2}g_{\m\n}, \\
&&\phi\rightarrow\Omega(x)^{-1}\phi,
\end{eqnarray}
however $S_{0}$ is not fully conformally invariant due to the presence of the usual 
$R$-term. Varying the first part of the action with respect to the metric provides the 
Einstein tensor, while varying the second term provides the energy-momentum 
tensor of a conformally-coupled scalar field, which in four dimensions reads
\be
T^{(\phi)}_{\m\n}=\frac{1}{2}\nabla_{\m}\phi\nabla_{\n}\phi-\frac{1}{4}g_{\m\n}\nabla_{\a}
\phi\nabla^{\a}\phi
+\frac{1}{12}\left(g_{\m\n}\square-\nabla_{\m}\nabla_{\n}+G_{\m\n}
\right)\phi^{2}~.
\ee

Let us now try to include higher-order terms in the action $S_0$, introducing 
additionally 
a new 
degree of freedom. For this shake, a second scalar $\Psi$ is added, allowing also for a 
non-minimally derivative coupling with the Einstein tensor, and moreover an additional 
coupling of this 
second 
scalar with the energy-momentum tensor of the conformally-coupled scalar $\phi$. Hence, 
we consider an additional action $S_{1}$ as
\be
S_1=\int dx^{4} \sqrt{-g}\; \left[\b G_{\m\n}\nabla^{\m}\Psi\nabla^{\n}\Psi-\g
T^{(\phi)}_{\m\n}\nabla^{\m}\Psi\nabla^{\n}\Psi\right] ~,
\label{johny}
\ee
where $\b$ and $\g$ are dimension-full coupling constants. Thus, in summary, we will 
consider the total action
\be 
\label{extended action}
 S=S_0+S_1.
 \ee

The reason for considering such an action is the following: The $\Psi$-field equation 
(obtained from (\ref{johny})), due to shift symmetry, can be nicely written as a current 
conservation equation,
namely
\be
\label{psii}
\mathcal{E}_{\Psi}=\nabla_{\m}J^{\m}=0~,
\ee
with
\be
\label{psiicurrent}
J^{\m}=-2\left(\b G^{\m\n}-\g T^{\m\n}
\right)\nabla_{\n}\Psi~. 
\ee
Therefore, it becomes obvious that the current vector $J^\mu$ ``encodes'' the metric 
field equations of the action (\ref{bek}) (this would become impossible if the sift 
symmetry was broken, for instance by including a general potential). That is why we can 
refer to the action 
$S_0$ as being the ``seed'' action, or the precursor, of the higher-order action $S_1$.

With this procedure, at the level of the action, an already existing theory can be 
extended by 
picking up its field equations and coupling it accordingly with an 
additional field. The resulting theory is now a bi-scalar tensor theory 
\cite{Padilla:2012dx,Ohashi:2015fma}, however the important advantage is that it is 
also partially conformally invariant \cite{Deser:1970zza,Padilla:2013jza}, and, similarly 
to the seed theory, it can also become fully conformally invariant if one 
neglects the first, Einstein-Hilbert term, and introduces either the Weyl derivative or 
properly 
rescaled fields \cite{Padilla:2013jza,Charmousis:2014zaa}. Furthermore, due to 
this 
construction method, the field equations are ensured that they will not contain more than 
two time 
derivatives,
which is an additional significant advantage.
One could follow the above 
procedure on 
and on, inserting additional scalars, resulting to partially conformally invariant 
multi-scalar theories. However in this work we desire to remain in the bi-scalar case for 
simplicity.

Variation of the action (\ref{extended action}) with respect to
the metric leads to
\bea
&&
\!\!\!\!\!
{\mathcal H}_{\m\n}=\frac{1}{16\pi G} G_{\m\n}-\eta \left[ 
\frac{1}{2}\nabla_{\m}\phi\nabla_{\n}\phi-\frac{1}{4}
g_{\m\n}\nabla_{\a}\phi\nabla^{\a}\phi+\frac{1}{12}\left(g_{\m\n}\square-\nabla_{\m}
\nabla_{\n}+G_{\m\n} \right)\phi^{2}\right]
\nn
\\
&&
\qquad  +\b \Big\{ 
-\frac{1}{2}g_{\m\n}G^{\a\b}\nabla_{\a}\Psi\nabla_{\b}\Psi+2G_{(\m}^{\,\,\,\,\,\la}\nabla_
{\n)}\Psi\nabla_{\la}\Psi+\frac{1}{2}R\nabla_{\m}\Psi\nabla_{\n}\Psi-\frac{1}{2} 
R_{\m\n}\nabla^{\a}\Psi\nabla_{\a}\Psi
\nn\\
&& 
\quad \quad 
\quad \ \ \ \
+\frac{1}{2}g_{\m\n}\left[(\square\Psi)^{2} 
-\nabla_{\a}\nabla_{\b}\Psi\nabla^{\a}\nabla^{\b}\Psi-R_{\a\b}\nabla^{\a}\Psi\nabla^{
\b}\Psi\right]\nn\\
&&\quad \quad \ \ \ \,\ \ \ \,
+\nabla_{\m}\nabla^{\a}\Psi\nabla_{\n}\nabla_{\a} 
\Psi-\square\Psi\nabla_{\m}\nabla_{\n}\Psi+R_{\m\,\,\,\n}^{\,\,\,\a\,\,\,\,\b}\nabla_{\a}
\Psi\nabla_{\b}\Psi\Big\}
\nn\\
&&\quad \quad 
+\g
\Bigg\{
\frac{1}{4}g_{\m\n}\nabla_{\a}\phi\nabla_{\b}\phi\nabla^{\a}\Psi\nabla^{\b}\Psi 
-\nabla_{(\m}\phi\nabla_{\n)}\Psi 
\nabla_{\a}\phi\nabla^{\a}\Psi-\frac{1}{8}g_{\m\n}\nabla_{\a}\phi\nabla^{\a}\phi\,\nabla_{
\b}\Psi\nabla^{\b}\Psi\nn\\
&&\quad 
\quad \quad \ \ \ 
+\frac{1}{4}\nabla_{\m}\phi\nabla_{\n}\phi\,\nabla_{\b}\Psi\nabla^{\b}\Psi+\frac{1}{4}
\nabla_{\a}\phi\nabla^{\a}\phi\,\nabla_{\m}\Psi\nabla_{\n}\Psi+\frac{1}{24}g_{ 
\m\n}\square(\phi^{2})\nabla_{\a}\Psi\nabla^{\a}\Psi\nn\\
&&\quad \quad \quad  \ \ \ 
-\frac{1}{12}\nabla_{\m}\nabla_{\n}
\phi^{2}\,\nabla_{\a}\Psi\nabla^{\a}\Psi
+\frac{1}{12}\nabla_{(\m}\left[\nabla_{\n)}(\phi^{2})\,\nabla_{\a}\Psi\nabla^{\a}\Psi 
\right]-\frac{1}{24}g_{\m\n}\nabla^{\b}(\nabla_{\b}\phi^{2}\,\nabla_{\a}\Psi\nabla^{\a}
\Psi)
\nn\\
&&
\quad \quad \quad \ \ \ 
-
\frac{1}{12}\square(\phi^{2})\nabla_{\m}\Psi\nabla_{\n}\Psi
-\frac{1}{24}g_{\m\n}\nabla_{\a}\nabla_{\b}\phi^{2}\,\nabla^{\a}\Psi\nabla^{\b}\Psi+\frac{
1}{6}\nabla_{\a}\nabla_{(\m}\phi^{2}\,\nabla_{\n)}\Psi\nabla^{\a}\Psi  
\nn\\
&&\quad \quad 
\quad\ \ \ 
-\frac{1}{12}\nabla_{\a}\left[\nabla_{(\m}\phi^{2}\,\nabla_{\n)}\Psi\nabla^{\a}
\Psi\right]
+\frac { 1 } {
24}\nabla_{\a} 
(\nabla^{\a}\phi^{2}\,\nabla_{\m}\Psi\nabla_{\n}\Psi)
+\frac{1}{24}\,\phi^{2}g_{\m\n}G^{
\a\b}\nabla_{\a}\Psi\nabla_{\b}\Psi
\nonumber \\
&&  
\quad \quad 
\quad\ \ \ 
-\frac{1}{6}
\phi^{2}G_{(\m}\,^{\la}\nabla_{\n)}\Psi\nabla_{\la}\Psi-\frac{1}{24}\phi^{2}R\nabla_{\m}
\Psi\nabla_{
\n}\Psi
+\frac{1}{24}\phi^{2}R_{\m\n}\nabla^{\a}\Psi\nabla_{\a}\Psi
\nonumber\\
&&
\quad \quad 
\quad\ \ \ 
+\frac{1}{12} 
\square\Psi\nabla_{(\m}\big(\phi^{2}\big)\nabla_{\n)}\Psi 
-\frac{1}{12}\nabla^{\a}\Psi\nabla_{(\m}\big(\phi^{2}\big)\nabla_{\n)}\nabla_{\a}
\Psi-\frac{1}{12}\nabla_{\a}\big(\phi^{2}\big)\nabla_{(\m}\nabla^{\a}\Psi\nabla_{\n)}
\Psi
\nonumber \\
&&
\quad \quad 
\quad\ \ \ 
+\frac{1}{
12}\nabla_{\a}\big(\phi^{2}\big)\nabla^{\a}\Psi\nabla_{\m}\nabla_{\n}\Psi
-\frac{1}{24}\square\big(\phi^{2}\big)\nabla_{\m}\Psi\nabla_{\n}\Psi-\frac{1}{24}
\nabla_{\m}\nabla_{\n}\big(\phi^{2}\big)\nabla^{\a}\Psi\nabla_{\a}\Psi
\nonumber
\\
&&
\quad \quad 
\quad \ \ \ 
-\frac{1}{24}g_{\m\n
}\nabla_{\a}\nabla_{\b}\big(\phi^{2}\big)\nabla^{\a}\Psi\nabla^{\b}\Psi
-\frac{1}{12}g_{\m\n}\nabla_{\a}\big(\phi^{2}\big)\nabla^{\a}\Psi\square\Psi+\frac{1}
{12}g_{\m\n}\nabla_{\a}\big(\phi^{2}\big)\nabla_{\b}\Psi\nabla^{\a}\nabla^{\b} 
\Psi
\nonumber
\ee
\bea
&&
\quad \ 
\quad\  
+\frac{1}{12}\nabla_{\a}
\nabla_{(\m}\big(\phi^{2}\big)\nabla^{\a}\Psi\nabla_{\n)}\Psi
+\frac{1}{
24}g_{\m\n}\square\big(\phi^{2}\big)\nabla^{\a}\Psi\nabla_{\a}\Psi
\nonumber
\\
&&
\quad \ 
\quad\   
-\frac{1}{12}\phi^{2}
\Big\{\frac{1}{2}g_{\m\n}
\Big[
(\square\Psi)^{2}-\nabla_{\a}
\nabla_{\b}\Psi\nabla^{\a}\nabla^{\b}\Psi-R_{\a\b}\nabla^{\a}\Psi\nabla^{\b} 
\Psi\Big]+\nabla_{\m}\nabla^{\a}\Psi\nabla_{\n}\nabla_{\a}\Psi 
\nonumber \\
&&\quad \
\quad \ \ \ \ \ \ \ \ \ \ \  \,
-\square\Psi\nabla_{\m}\nabla_{\n}\Psi+R_{\m}\,^{\a}\,_{\n}\,^{\b}\nabla_{\a}
\Psi\nabla_{\b}\Psi\Big\}
\Bigg\}=0~,
\label{metriceom0}
\ee
where the parentheses in space-time indices denote symmetrization.

Furthermore, variation of (\ref{extended action}) with respect to $\phi$ leads to
\bea
&&
\!
{\mathcal E}_{\phi}=
\eta \left(\square \phi-\frac{1}{6}R \phi \right)\nn \\
&&
\quad \quad 
+\g 
\left[\frac{5}{6}\nabla_{\n}\left(\nabla^{\n}\Psi\nabla_{\m}\phi\nabla^{\m}\Psi 
\right)-\frac{1}{3}\nabla^{\a}\left(\nabla_{\a}\phi\nabla_{\m}\Psi\nabla^{\m}\Psi 
\right)-\frac{1}{6}
\square \phi 
\nabla_{\m}\Psi\nabla^{\m}\Psi\right.
\nn \\
&&
\quad \quad \quad \ \ \
+\frac{1}{6}\nabla_{\m}\nabla_{\n}\phi\nabla^{\m}\Psi\nabla^
{\n}\Psi
-\frac{1}{6}\phi\, 
G_{\m\n}\nabla^{\m}\Psi\nabla^{\n}\Psi+\frac{1}{6}\square\Psi\,\nabla_{\n}\phi\nabla^{\n}
\Psi+\frac{1}{6}\phi\left( 
\square\Psi\right)^{2}\nn 
\\
&&
\left.
\quad \quad \quad
\ \ \
-\frac{1}{6}\nabla_{\n}\phi\nabla^{\m}\Psi\nabla_{\m}\nabla^{\n}
\Psi
-\frac{1}{6}\phi\,\nabla_{\m}\nabla_{\n}\Psi\nabla^{\m}\nabla^{\n}\Psi-\frac{1}{6}\phi\,R_
{\m\n}\nabla^{\m}\Psi\nabla^{\n}\Psi
\right]=0.
\label{phieom0}
\eea
Finally, as we have already mentioned, variation of the action (\ref{extended action}) 
with respect to $\Psi$ leads to equation (\ref{psii}), which is its equation of motion.

\section{Cosmology}
\label{Cosmolappl}

In the previous section we demonstrated how to built a bi-scalar theory that maintains 
the 
conformal 
invariance of the ``seed'' single-scalar theory (up to the Einstein-Hilbert term). In 
this section we are interested in applying this theory at a cosmological framework. 
Hence, we consider a flat Friedmann-Robertson-Walker (FRW) spacetime metric of the form
 \be
ds^{2}=-dt^{2}+a(t)^{2} \delta_{ij}dx^{i}dx^{j},
\ee
with $a(t)$ the scale factor.
Additionally, we consider the matter sector, described by a perfect-fluid action $S_{m}$, 
and thus the total action will be
\be
S=S_{0}+S_{1}+S_{m}.
\ee
In this case, variation with respect to the metric leads to the metric-field equations
\be
\mathcal{H}_{\m\n}=\frac{1}{2}T^{(m)}_{\m\n},
\ee
with $\mathcal{H}_{\m\n}$ given by (\ref{metriceom0}) and 
where $T^{(m)}_{\m\n}=\frac{-2}{\sqrt{-g}}\frac{\d S_{m}}{\d g^{\m\n}}$ is the 
energy-momentum tensor of the matter perfect fluid. In particular, in the case of FRW 
geometry, 
they write explicitly as
\begin{equation}
\frac{3}{8\pi G}
H^{2}-\rho_m-\frac{\eta}{2}\left(H\f+\dot{\f}\right)^{2}
-9\b H^{2}\dot{
\Psi}
^{2}+\frac{3\gamma }{4} \left(H\f+\dot{\f}\right)^{2}\dot{\Psi}^{2}=0,
  \label{FR1}
\end{equation}
\bea
&&
\!\!\!\!\!\!\!\!\!\!\!\!\!\!\!\!
\frac{1}{8\pi G}\left(3H^{2}+2\dot{H}\right)+p_m
+\frac{\eta}{6}
\left[\dot{\f}^{2}-3H^{2}\f^{2}
-4H\f\dot { \f}-2\f\left(\f 
\dot{H}+\ddot{\f}\right)\right]
\nn\\
&& \ \ 
-\b\dot{\Psi}
\left[\left(3H^{2}+2\dot{
H}\right)\dot{\Psi}+4H\ddot{\Psi}\right]
\nn\\
&&\ \ 
+\frac{\gamma}{12}
\dot{\Psi}\left\{\dot{\Psi}\left[3H^{
2}\f^{2}+4H\f\dot{\f}-\dot{\f}^{2}+2\f\left(\f\dot{H}+\ddot{\f}
\right)\right]+4\f\left(H\f+\dot{\f}\right)\ddot{\Psi}\right\}=0,
  \label{FR2}
\eea
which are just the Friedmann equations of the scenario at hand. In the above expressions
dots  denote differentiation with respect to $t$,  $H=\dot{a}/a$ is the Hubble 
parameter, and $\rho_m$ and $p_m$ are respectively the matter energy density and pressure.
Similarly, the two scalar field equations (\ref{phieom0}) and  (\ref{psii}) in the case 
of FRW geometry become:
\be
\mathcal{E}_{\phi}=-\frac{1}{2}\left(2\eta-\gamma\dot{\Psi}^{2}\right)\left[\f\left(2H^{2}
+\dot{H}\right)+3H\dot{\f}+\ddot{\f}\right]+\gamma\left(H\f+\dot{\f}\right)\dot{\Psi}\ddot
{\Psi}=0,
\label{phieom2}
\ee
and
\bea
&&\!\!\!\!\!\!\!\!\!\!\!\!\!\!
\mathcal{E}_{\Psi}=6\b 
H\left[\left(3H^{2}+2\dot{H}\right)\dot{\Psi}+H\ddot{\Psi}\right]\nonumber\\
&&
-\frac{1}{2}
\gamma\left(H\f+\dot{\f}\right)
\left\{\dot{\Psi}
\left[3H^{2}\f+5H\dot{\f}+2\left(\f\dot{H}
+\ddot{\f}\right)\right]+\left(H \f+\dot{\f}\right)\ddot{\Psi}\right\}=0.
\label{psii2}
\eea
Finally, one can straightforwardly  
verify that the above equations satisfy the equations 
arising from the fact that the total action is diffeomorphism  invariant 
\cite{Ohashi:2015fma}:
\be
\nabla_{\m}\mathcal{H}^{\m\n}+\frac{1}{2}\mathcal{E}_{\f}\nabla^{\n}\f+\frac{1}{2}\mathcal
{E}_{\Psi}
\nabla^{\n}\Psi=\frac{1}{2}\nabla_{\m}T^{\m\n}=0.
\ee
Lastly, we stress that the above equations do not contain higher than two 
time-derivatives, as expected due to the construction method we followed in order to 
build this conformal bi-scalar scenario.

Concerning the late-time application of the above equations, we can immediately see that 
we can re-write the two Friedmann equations 
(\ref{FR1}),(\ref{FR2})  in the usual form, namely
\begin{eqnarray}
&&H^2=\frac{8\pi G}{3}(\rho_{DE}+\rho_m)
 \label{FR1b}
 \\
&&2\dot{H}+3H^2=-8\pi G(p_{DE}+p_m),
 \label{FR2b}
\end{eqnarray}
if we define an effective dark energy sector with energy density and pressure 
respectively as:
\begin{eqnarray}
  \label{rhoDE}
&& \!\!\!\!\!\!\!\!\!\!\!\!\!\! \!\!\!\!\!\!\!\!\!\!\!\!\!\! \!\!\!\!\!\!\!\!\!\!\!\!\!\! 
\!\!\!\!\!\!\!\!\!\!\!\!\!\! \!\!\!\!\!\!\!\!\!\!\!\!\!\! \!\!\!\!\!\!\!
 \rho_{DE}\equiv 
 \frac{\eta}{2}\left(H\f+\dot{\f}\right)^{2}
+9\b H^{2}\dot{
\Psi}
^{2}-\frac{3\gamma }{4}\left(H\f+\dot{\f}\right)^{2}\dot{\Psi}^{2}  ,
 \end{eqnarray}
\begin{eqnarray}
\label{pDE}
&& \!\!\!\!\!\!\!\!\!\!\!\!\!\! 
p_{DE}\equiv 
\frac{\eta}{6}
\left[\dot{\f}^{2}-3H^{2}\f^{2}
-4H\f\dot { \f}-2\f\left(\f 
\dot{H}+\ddot{\f}\right)\right]
\nn\\
&& \ \ 
- \b\dot{\Psi}
\left[\left(3H^{2}+2\dot{
H}\right)\dot{\Psi}+4H\ddot{\Psi}\right]
\nn\\
&&\ \ 
+\frac{\gamma}{12}
\dot{\Psi}\left\{\dot{\Psi}\left[3H^{
2}\f^{2}+4H\f\dot{\f}-\dot{\f}^{2}+2\f\left(\f\dot{H}+\ddot{\f}
\right)\right]+4\f\left(H\f+\dot{\f}\right)\ddot{\Psi}\right\}  .
\end{eqnarray}
Thus, in the scenario at hand  we acquire an 
effective dark-energy sector that consists of both scalar fields. Additionally,  
using their   equations of motion (\ref{phieom2}) and (\ref{psii2}), we can verify that
\begin{equation}
\dot{\rho}_{DE}+3H(\rho_{DE}+p_{DE})=0,
\end{equation}
and we can define the equation-of-state parameter for the effective dark-energy sector as
\begin{equation}
\label{wDE00}
w_{DE}\equiv \frac{p_{DE}}{\rho_{DE}}.
\end{equation}
Lastly, we mention that as usual the matter energy density and pressure satisfy the 
 equation 
\begin{equation}
\label{rhoevol00}
\dot{\rho}_m+3H(\rho_m+p_m)=0.
\end{equation}

Before proceeding, it is worthy to mention that the equations of motion for the 
scalar fields $\f$ and $\Psi$ admit a first integral. In particular, (\ref{phieom2}) can 
be 
written as
\bea
&&2a^{2}\mathcal{E}_{\f}=-\partial_{t}\left[\left(2 \eta-\g \dot{\Psi}^{2} \right) a 
\left(\phi \dot{a}+a \dot{\phi} \right) \right],
\eea
while 
(\ref{psii2})  can be 
written as
\bea
&&a^{3}\mathcal{E}_{\Psi}=\partial_{t}\left[ a^{3}
 J^{0} \right]=0 
 \label{inegr2help}
\eea
with 
\be
J_0=\left[\left(-12\b+\g\phi^{2}\right)\dot{a}^{2}+2\g\,a\,\f\,\dot{a}\dot{\f}+\g\,a^{2}\,
\dot{\f}^{
2}\right]\,\dot{\Psi},
\ee
 and hence we respectively obtain:
\bea
&&\left(2 \eta-\g \dot{\Psi}^{2} \right) a \left(\phi \dot{a}+a \dot{\phi} \right)=c_{0}
\label{integral1}
\\
&&-6\b a \dot{a}^{2}\dot{\Psi}+\frac{1}{2}\g a \left(\f \dot{a}+a \dot{\f} 
\right)^{2}\Dot{\Psi}=c_{
1},
\label{integral2}
\eea
where $c_{0},c_{1}$ are integration constants.
These expressions are going to be crucial in the following section, since they allow us
to extract analytical solutions.

\section{Solutions}
\label{SolutionsSec}

In the previous section we derived the cosmological equations in a bi-scalar model which 
exhibits partial conformal invariance. In this section we 
are first interested in extracting analytical solutions  in the case where the matter 
sector is absent, and moreover to investigate the scenario numerically in the case where 
matter is present. As we mentioned above, the equations of motion for the scalar fields 
admit the first integrals (\ref{integral1}),(\ref{integral2}), which proves to be crucial 
in the solution extraction.

\subsection{Case 1: $c_0=0$}

Let us first neglect the matter sector and investigate the case where the first integral 
of the $\phi$-field equation, i.e. (\ref{integral1}), is equal to zero, namely we 
consider $c_0=0$. Hence, we acquire two subcases, corresponding to which of the two 
parentheses terms becomes zero.

\begin{itemize}

\item

In the first subcase, namely when the first parenthesis in (\ref{integral1}) is zero,
we have that
\be
2 \eta-\g \dot{\Psi}^{2}=0,
\ee
which admits the solution 
\be
 \Psi(t)=\pm\sqrt{\frac{2\eta}{\g}}\,\,t+\Psi_{0},
 \label{Psisolcase1}
\ee
with $\Psi_{0}$ an integration constant. Both sign-branches lead to the same 
observable results, since both the Friedmann equations (\ref{FR1}),(\ref{FR2}), as well 
as the scalar-field equations (\ref{phieom2}), (\ref{psii2}) depend only on 
$\dot{\Psi}^2$ and/or $\dot{\Psi}\ddot{\Psi}$. Substituting the above solution 
(\ref{Psisolcase1}) for $\Psi(t)$ into the second Friedmann equation (\ref{FR2})
we obtain
\be
\label{c0=0firstcase}
\left(\dot{a}^{2}+2a\ddot{a}\right)(\g-16 \pi G \b\eta)=0.
\ee

If the first bracket of (\ref{c0=0firstcase}) is zero, namely if 
$\dot{a}^{2}+2a\ddot{a}=0$, then we immediately extract the solution 
\be
 a(t)=a_0\left(t-t_{0}\right)^{2/3},
\ee
where $a_{0}$, $t_0$ are integrations constant. Thus, substituting the above $\Psi(t)$ 
and $a(t)$ into the first Friedmann equation (\ref{FR1}) we finally acquire
the solution for $\phi$ as
\be
\phi(t)=\frac{\f_0}{(t-t_0)^{2}}\pm\frac{\sqrt{6}\sqrt{3\b\eta-\g/16 \pi 
G}}{\sqrt{\g\eta}},
\ee
with $\f_{0}$   an integration constant. Note that the constant $t_0$ can be set to 
zero without loss of generality.

Interestingly enough, we observe that in the above solution the universe behaves as a 
matter-era  although we have not explicitly considered the matter sector. This can be 
immediately explained, since in this case the ``effective'' dark energy sector 
constituted from the two scalar fields obtains an equation-of-state parameter equal to 
zero, as can be seen from (\ref{wDE00}). Hence, in the scenario at hand we have obtained 
a 
form of ``mimetic'' dark matter \cite{Chamseddine:2013kea,Leon:2014yua}, constituted from 
the two scalar fields. This important feature deserves further investigation, in 
particular examining the behavior of this solution under perturbations and confronting it 
with observational data from large-scale structure \cite{Bernardeau:2001qr}. Such a 
project is left for a future investigation.

If now the second bracket of (\ref{c0=0firstcase}) is zero, namely $\g-16 \pi G\b\eta=0$, 
then 
the first Friedmann equation (\ref{FR1}) gives
\be
\f(t)=\frac{\phi_0}{a(t)}\pm2\sqrt{\frac{3 \b}{\g}}.
\label{phiasol}
\ee
Substituting this solution into  the field equation (\ref{psii2}) for $\Psi(t)$ we see 
that it is immediately satisfied, which was expected since in this case  $J^{0}$ in 
 (\ref{inegr2help}) is identically zero. Hence, this solution branch is compatible with 
every cosmological evolution $a(t)$. This is a important, since one can 
obtain arbitrary scale-factor behavior with only tuning the parameters as required for 
this branch, namely $c_0=0$ and $\g-16 \pi G\b\eta=0$. This feature is a significant 
advantage of the scenario at hand, since apart from late-time acceleration one 
can describe the complete thermal 
history of the universe, namely the sequence from inflation, to radiation, to matter 
and to dark-energy epochs, as well as alternatively evolutions such as the bouncing or 
cyclic ones \cite{Novello:2008ra}.

\item

In the second subcase, namely when the second parenthesis in (\ref{integral1}) is zero, 
we have that
\be
\phi \dot{a}+a \dot{\phi}=0,
\label{phia1}
\ee
which leads to the solution
\be
  \f(t)=\frac{\phi_0}{a(t)}.
  \label{phia1bb}
\ee
Substituting into the first Friedmann equation  (\ref{FR1}) we obtain
\be
\Psi(t)=\pm\sqrt{ \frac{1}{24 \pi G\b}}\,\, t+\Psi_{0},
\label{Psilinear0}
\ee
with $\Psi_{0}$ an integration constant. Moreover, inserting these into the second 
Friedmann equation  (\ref{FR2}) we acquire $\dot{a}^{2}+2a\ddot{a}=0$, which admits the 
solution
\be
 a(t)=a_0\left(t-t_{0}\right)^{2/3}.
\ee
Finally, substituting this into relation (\ref{phia1bb}) we obtain:
\be
  \f(t)=\frac{\phi_0}{a_0\left(t-t_{0}\right)^{2/3}}.
\ee
Similarly to the previous subcase, we can see that we obtain a matter era despite the 
fact that we have not considered an explicit matter sector, namely we obtain a mimetic 
dark matter arising from the combination of the two coupled scalar fields.
This is a great advantage of the scenario at hand.

 \end{itemize}

\subsection{Case 2: $16 \pi G\eta \b-\g=0$}

Let us now investigate another class of solutions, characterized by the   parameter 
relation  $16 \pi G\eta \b-\g=0$ (note that in contrast with the second branch of the 
first 
subcase of the previous subsection we keep a general $c_0$), still without considering 
an explicit matter sector. 
In this case the first Friedmann equation (\ref{FR1}) acquires the factorized form
\be
\left(3\b\dot{\Psi}^{2}-\frac{1}{8 \pi G}\right)\left[12\b 
\dot{a}^{2}-\g\left(\f\dot{a}+a\dot{\f}\right)^{2}\right]=0.
\label{eqcase2A1}
\ee

\begin{itemize}

\item  

If the first bracket of (\ref{eqcase2A1}) is zero then $\Psi(t)$ obtains the linear 
relation (\ref{Psilinear0}), namely
 \be
 \Psi(t)=\pm\sqrt{\frac{1}{24 \pi G\b}}t+\Psi_{0},
\ee
 however in this case the first integral (\ref{integral1}) leads to
\be
\f(t)=\frac{1}{a(t)}\left[\f_{
0}+\frac{12   \pi G c_{0}\b}{\g }  \int\frac{dt'}{ a(t')} \right].
\ee
Hence, substituting these into the second Friedmann equation (\ref{FR2})
 we can easily obtain  
\be
H^{2}=\frac{12 \pi^2 G^2 c_{0}^{2}\b}{ \g  a^{4}}+\frac{C_{1}}{a^{3}},
\label{FrDMDE}
\ee
where $C_{1}$ is a new integration constant. Interestingly enough, we observe that in 
this subclass of solutions we obtain in the Friedmann equation both an effective dark 
matter sector, as well as an effective dark radiation sector 
\cite{Mukohyama:1999qx,Saridakis:2009bv}, despite the fact 
that we have not considered explicitly such sectors. This is a great advantage of the
scenario, since the two coupled scalar fields can give rise to these 
effective sectors in a mimetic way. This feature could have very interesting physical 
implications. 

The solution of (\ref{FrDMDE}) is straightforward, it is a fractional function of $t$, 
and since it exists in the literature \cite{Peebles:1994xt} we do not write it 
explicitly here. We 
just mention that in the case where $c_0=0$ , i.e. when the effective dark 
radiation disappears, we obtain the usual matter era evolution, namely $ 
a(t)=\left(3/2\right)^{2/3} C_1^{1/3}\left(t-t_{0}\right)^{2/3}$, while in the case where 
$C_1=0$, i.e. when the effective dark matter disappears, we obtain the usual radiation 
era evolution, namely  
 $a(t)=\sqrt{2} \left(\frac{12 \pi^2 G^2  c_{0}^{2}\b}{ \g 
}\right)^{1/4}\left(t-t_{0}\right)^{1/2}$. However, in general, using (\ref{FrDMDE}) one 
can reconstruct the observed thermal history of the universe,  with an initial radiation 
era (for small scale factors) followed by the matter epoch (for larger scale factors).

\item 

If now the second bracket of (\ref{eqcase2A1}) is zero then $\phi(t)$ obtains the  
solution (\ref{phiasol}), namely
\be
\f(t)=\frac{\phi_0}{a(t)}\pm2\sqrt{\frac{3\b}{\g}},
\ee
however in this case the first integral (\ref{integral1}) leads to
\be
\Psi(t)=\int\sqrt{\frac{1}{ 8\pi G  \b}-\frac{c_{0}}{2\sqrt{3\b\g}\,a\,\dot{a}}} 
dt+\Psi_{0}.
\ee
Note that the above two solutions of $\f(t)$ and $\Psi(t)$ with 
respect to $a(t)$ satisfy all the field equations for arbitrary scale factor. Hence, this 
solution subclass can reproduce any cosmological evolution, which is a significant 
advantage of the scenario at hand, revealing its capabilities.

 \end{itemize}

\subsection{General case}

Let us now investigate the general solution subclass without the presence of an explicit 
matter sector. The cosmological system consists of the two Friedmann equations 
(\ref{FR1}),(\ref{FR2}) and the two scalar field equations expressed as integrals, namely 
(\ref{integral1}),(\ref{integral2}), with only three of them being independent.
Considering for $\f(t)$ the ansatz 
\be
\f(t)=\frac{k(t)}{a(t)},
\ee 
then (\ref{FR1}),(\ref{integral1}),(\ref{integral2}) respectively become
\bea
\dot{a}^{2}\left(36 \b \dot{\Psi}^{2}-\frac{3}{2\pi G} 
\right)+\dot{k}^{2}\left(2\eta-3\g\dot{\Psi}^{2}\right)&
=&0\label{meq1}\\
a \dot{k}\left(2\eta-\g \dot{\Psi}^{2}\right)-c_{0}&=&0\label{meq2}\\
\frac{1}{2}a\left(\g\dot{k}^{2}-12\b\dot{a}^{2}\right)\dot{\Psi}-c_{1}&=&0.\label{meq3}
\eea

\begin{itemize}

\item 

In the trivial case where   $36 \b \dot{\Psi}^{2}-\frac{3}{2\pi G}=0$, which implies that 
$
\Psi(t)=\sqrt{\frac{1}{24 \pi G \b}}t+\Psi_{0}$,
Eq.  (\ref{meq1}) leads either to  $16 \pi G \eta\b-\g=0$, case which was examined in the 
previous subsection,  or to  $\dot{k}=0$, which leads to 
\be
\f(t)=\frac{\f_0}{a(t)}.
\ee 
Substituting in (\ref{meq2}),(\ref{meq3}) we see that 
the full system is 
satisfied if
\be
a(t)=a_0( t-t_{0})^{2/3},
\ee 
with the additional parameter constraint 
$9  \sqrt{\pi G} \, c1+2\sqrt{6}\,a_{0}^{3}\,\sqrt{\b}=0$. Once again we observe that we 
obtain a matter 
era, although we have not considered an explicit matter sector, due to the fact that the 
two scalars produce an effective matter sector of mimetic nature.

\item  

In the general case where $36 \b \dot{\Psi}^{2}-\frac{3}{2 \pi G}\neq0$,  Eq.  
(\ref{meq1}) leads to
\be
\label{meq1b}
\dot{a}=\pm\dot{k}\sqrt{\frac{\left(
3\g\dot{\Psi}^{2}-2\eta\right)}{ 
\left(36 \b \dot{\Psi}^{2}-\frac{3}{2 \pi G} \right)}},
\ee
where we have assumed that $\dot{\Psi}^{2}$ lies in the appropriate 
ranges for the above relation to hold, namely
\bea
& 
\frac{3}{72 \pi G \b}<\dot{\Psi}^{2}\leq\frac{2\eta}{3\g}\quad\,\,\,\text{if} 
\quad\,\,\,
\frac{2\eta}{\g}<\frac{1}{8 \pi G \b}&\\
&  \frac{2\eta}{3\g}\leq 
\dot{\Psi}^{2}< 
\frac{3}{72 \pi G \b}\quad\,\,\,\text{if} \quad\,\,\,  
\frac{2\eta}{\g}>\frac{1}{8 \pi G \b}&.
\eea
Substituting (\ref{meq1b}) into  (\ref{meq3}) 
we obtain
\be
 (\g-16 \pi G \b\eta)a\,\dot{k}^{2}\dot{\Psi}=2c_{1}\,(1-24  \pi G   \b\dot{\Psi}^{2}).
 \label{auxeq1}
\ee
Hence, disregarding the case where $16 \pi G  \eta\b-\g=0$ analyzed in the previous 
subsection, 
or the trivial cases $\dot{\Psi}=0$ which leads to an effective radiation era of the 
form $a(t)=a_0( t-t_{0})^{1/2}$, and $\dot{k}=0$ which leads to an effective matter era 
of the form  $a(t)=a_0( t-t_{0})^{2/3}$, Eq. (\ref{auxeq1}) leads to 
\be
\label{meq3b}
a(t)=\frac{2c_{1}\,(1-24  \pi G   \b\dot{\Psi}^{2})}{(\g-16 \pi G 
\b\eta)\dot{k}^{2}\dot{\Psi}}.
\ee
Inserting this expression into (\ref{meq2}) we acquire
\be
\label{meq2b}
\dot{k}=\frac{2c_{1}\left(1-24 \pi G  \b\dot{\Psi}^{2}\right) 
\left(2\eta-\g\dot{\Psi}^{2}\right)}{c_{0}\left(\g-16  \pi G \b\eta \right)\dot{\Psi}},
\ee
and finally combing (\ref{meq1b}), with (\ref{meq3b}) and (\ref{meq2b}) we arrive on 
a master equation
\bea
&&
\frac{c_{0}^{4}\left(\frac{\g}{16  \pi G}-\b\eta\right)^{2}
\left[\frac{\eta}{4  \pi G }+6\left(\frac{\g}{16  \pi 
G}+\b\eta\right)\dot{\Psi}^{2}-15\b\g\dot{\Psi}^{4} 
\right]^{2}\ddot{\Psi}^{2}}{c_{1}^{2} (\frac{1}{8 \pi G}-3\b\dot{\Psi}^{2})^{4}
 (\g\dot{\Psi}^{2}-2\eta)^{6}}
\nonumber\\
&&
+\frac{c_{1}^{2} (\frac{1}{8 \pi G}-3\b\dot{\Psi}^{2}) (\g\dot{\Psi}^{2}-2\eta) 
(3\g\dot{\Psi}^{2}-2\eta)}{12c_{0}
^{2}(\frac{\g}{16  \pi G}-\b\eta)^{2}\dot{\Psi}^{2}}
=0.
\label{mastereq1}
\eea

The solution for ${\Psi}(t)$ of this equation allows us to go back and extract the 
solutions for the remaining involved quantities, namely $a(t)$ and $\phi(t)$. 
Unfortunately, the above equations cannot be solved analytically in general. Hence, one 
should either investigate the subcases that we have already studied in the previous 
subsections, or solve the full equation approximately expanding it and keeping the lower 
orders, or solve it numerically. However, since in the following subsection we will 
elaborate numerically  the full cosmological equations, namely including the explicit 
matter sector, we are not interested in solving numerically (\ref{mastereq1}), which 
corresponds to the subclass without matter.

\end{itemize}

\subsection{Numerical elaboration}

In the previous subsections we presented analytical solutions of the cosmological system 
at hand, in the case where an explicit matter sector is absent. In this subsection we 
will investigate the full cosmological equations, including the matter sector, namely 
equations (\ref{FR1}),(\ref{FR2}),(\ref{phieom2}), (\ref{psii2}), focusing at late 
times. Since these equations cannot be solved analytically, we will perform  a numerical 
elaboration. 

As usual, we assume that the explicit matter sector corresponds to a dust fluid, namely 
we consider $p_m\approx0$. Furthermore, in order for our cosmological evolution to be 
consistent with observations, we impose the present values of the density parameters to be
$\Omega_{m0}=\frac{8\pi G\rho_{m0}}{3H^2}\approx0.3$ and
$\Omega_{DE0}=\frac{8\pi G\rho_{DE0}}{3H^2}\approx0.7$ \cite{Ade:2015xua}. Finally, we use 
 the redshift $z=-1+a_0/a$ as the independent variable, setting 
the current scale factor  $a_0$ to 1.

\begin{figure}[ht]
\begin{center}
\includegraphics[width=0.8\textwidth]{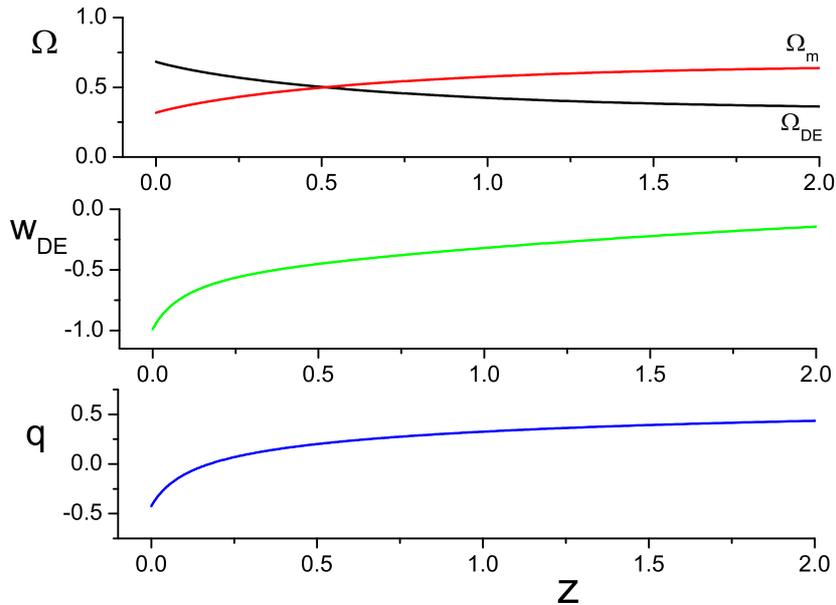}
\caption{{\it{The late-time cosmological evolution, for
the parameters choice $\eta=1$, $\beta=-11$, $\gamma=-1$ in units where $8\pi G=1$, 
having 
imposed 
$\Omega_{m0}\approx0.3$,
$\Omega_{DE0}\approx0.7$ at present, and having set the present
scale factor $a_0=1$. We use the redshift
$z=-1+a_0/a$ as independent variable. In the upper graph we present the 
evolution of the dark energy and matter density parameters. In the middle graph we
depict the evolution of the dark-energy equation of state. In the lower graph we present 
the evolution of
the deceleration parameter $q$.}}} \label{DEfig}
\end{center}
\end{figure}

In Fig. \ref{DEfig} we show the cosmological evolution for the parameters choice 
$\eta=1$, 
$\beta=-11$, $\gamma=-1$ in units where $8\pi G=1$, focusing on various 
observables. Specifically, in the upper graph we present the evolution of the matter and 
dark energy density parameters, defined as $\Omega_i=8\pi G\rho_i/(3 H^2)$, and as we can 
see it is in agreement with the observed one \cite{Ade:2015xua}. In the middle graph of 
Fig. 
\ref{DEfig} we depict the behavior of the dark-energy equation-of-state
parameter $w_{DE}$, which presents a dynamical behavior, acquiring at present a value 
very close to the cosmological-constant one, as expected from observations. Finally, in 
the lower graph of Fig. \ref{DEfig} we present the evolution of the deceleration 
parameter 
$q=-1-\dot{H}/H^2$. As we observe, the universe passed from deceleration ($q>0$) to 
acceleration ($q<0$) in the recent cosmological past, as it is required from 
observations.
In summary, the examined scenario of bi-scalar gravity can lead to a cosmological 
behavior in agreement with observations. We mention that we have not considered neither 
an explicit cosmological constant nor a potential that could play the role of a 
cosmological constant in particular limits, and therefore the late-time acceleration is 
a pure result 
of the novel, bi-scalar construction and the involved couplings between the two scalars.

\begin{figure}[ht]
\begin{center}
\includegraphics[width=0.85\textwidth]{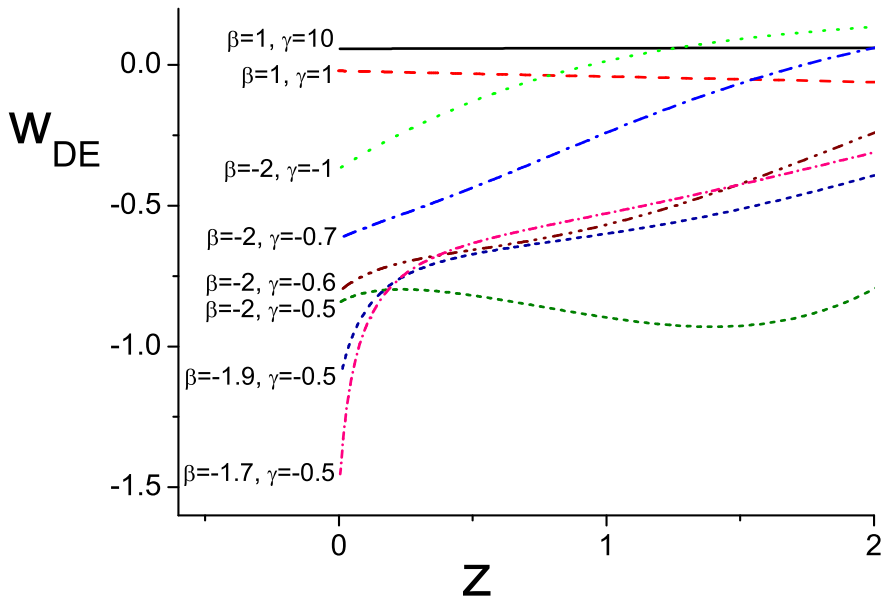}
\caption{{\it{The evolution of the dark-energy equation-of-state parameter as a 
function of the redshift $z=-1+a_0/a$, for $\eta=1$, and for eight choices of the 
parameters $\beta$  and $\gamma$, in units where $8\pi G=1$, having imposed 
$\Omega_{m0}\approx0.3$,$\Omega_{DE0}\approx0.7$ at present, and having set the present
scale factor $a_0=1$.  }}} \label{weffectfig}
\end{center}
\end{figure}

We now proceed to the investigation of how the two-scalar coupling parameters $\beta$ and 
$\gamma$ in action $S_1$ affect the cosmological evolution, and in particular the 
dark-energy equation-of-state parameter $w_{DE}$.  In Fig. \ref{weffectfig} we depict 
the evolution of $w_{DE}$ for various values of  $\beta$ and $\gamma$. As we can see, 
there are parameter regions for which  $w_{DE}$ lies in the quintessence regime, however 
there are also parameter regions for which  $w_{DE}$ exhibits the phantom-divide crossing 
in the recent cosmological past, acquiring a value below the cosmological-constant one at 
present. This is a significant advantage of the present bi-scalar scenario, since the 
phantom behavior is acquired although the whole construction is  free of ghosts and 
the two scalar fields are not phantom ones.

\section{Conclusions}
\label{Conclusions}

In this work we investigated the cosmological applications of a recently proposed 
gravitational modification, corresponding to a bi-scalar theory constructed in a way 
to exhibit partial conformal invariance, which could become full conformal invariance 
in the absence of the usual Einstein-Hilbert term and introducing additionally either 
the Weyl derivative or properly rescaled fields \cite{Charmousis:2014zaa,Padilla:2013jza}.
Such a theory is constructed by considering the action of a non-minimally 
conformally coupled scalar field as a ``seed'' action, in which one adds a second scalar 
allowing for a  nonminimal derivative coupling  with the Einstein tensor and the 
energy-momentum tensor of the first conformally-coupled scalar field. In this way, the 
equation of motion of the second field can be written as a current conservation equation, 
with the current enclosing the metric field equations of the initial ``seed'' 
action.

Applying this bi-scalar modified gravity in a cosmological framework, we extracted the 
Friedmann equations as well as evolution equations of the two scalar fields, obtaining an 
effective dark-energy sector constituted from both scalars. Firstly, we extracted various 
analytical solutions in the absence of an explicit matter sector. Interestingly enough, 
for some specific solution subclasses we saw that we can obtain a universe dynamics 
corresponding to matter-era evolution, although we have not considered an explicit matter 
sector. This important feature results from the fact that for this parameter region the 
effective dark-energy equation-of-state parameter becomes zero, and thus the scalar 
fields give rise not to an effective-dark energy sector, but to an affective dark-matter 
one, i.e. to a form of ``mimetic dark matter''. Additionally, for different parameter 
regions we showed that we can obtain both an effective dark-matter sector, as well as an 
effective dark-radiation sector, and hence obtaining the thermal history of the universe, 
with an initial radiation era  followed by the matter epoch. Finally, there are parameter 
regions that allow for an arbitrary scale-factor evolution, which can have important 
implications, since apart from late-time acceleration they can describe the complete 
thermal history of the universe, namely the sequence from inflation, to radiation, to 
matter and to dark-energy epochs, as well as alternatively evolutions such as the 
bouncing 
or cyclic ones.

In the case where an explicit matter sector is included, we evolved the full cosmological 
system numerically, focusing on various observables such as the matter and 
dark-energy density parameters, the deceleration parameter, and the effective dark energy 
equation of state. As we saw, the obtained cosmological evolution is in agreement with 
observations, with the matter era followed by the dark-energy epoch and the transition to 
the cosmic acceleration. Furthermore, for particular regions of the model parameters, the 
effective dark-energy equation-of-state parameter can pass through the cosmological 
constant value, resulting in the phantom regime at present. This feature reveals the 
capabilities of the (partially) conformally invariant bi-scalar theory, since 
the phantom behavior is acquired although the fields are canonical and the theory is 
ghost free. We stress here that the above behaviors are obtained without the presence 
of an explicit cosmological constant, or of a potential that could play the role of a 
cosmological constant in particular limits, i.e. they arise purely from the novel, 
bi-scalar construction and the involved couplings between the two fields.

In summary, (partially) conformally invariant bi-scalar theories have interesting 
cosmological implications in agreement with observations. Thus, it would be worthy to 
further investigate them, confronting them with  observations using Type Ia 
Supernovae (SNIa), Baryon Acoustic Oscillations (BAO), and Cosmic Microwave Background 
(CMB) data, as well as analyzing the perturbations and constraining them with  
large-scale 
structure observations. Additionally, one could perform a detailed dynamical-system 
analysis, in order to by-pass the non-linearities and reveal the asymptotic 
cosmological behavior. These studies are left for  future projects.

\begin{acknowledgments}
MT is supported by TUBITAK 2216 fellowship under the application number 1059B161500790.
\end{acknowledgments}

\end{document}